# 1/$f$ Noise In Low Density Two-Dimensional Hole Systems In GaAs

G. Deville[*], R. Leturcq[*], D. L'Hôte[*], R. Tourbot[*], C.J. Mellor[†], M. Henini[†]

[*]*Service de Physique de l'Etat Condensé, DSM, CEA- Saclay, F-91191 Gif-sur-Yvette Cedex, France.*
[†]*School of Physics and Astronomy, University of Nottingham, University Park, Nottingham NG7 2RD, United Kingdom.*

**Abstract.** Two-dimensional electron or hole systems in semiconductors offer the unique opportunity to investigate the physics of strongly interacting fermions. We have measured the 1/$f$ resistance noise of two-dimensional hole systems in high mobility GaAs quantum wells, at densities below that of the metal-insulator transition (MIT) at zero magnetic field. Two techniques voltage and current fluctuations were used. The normalized noise power $S_R/R^2$ increases strongly when the hole density or the temperature are decreased. The temperature dependence is steeper at the lowest densities. This contradicts the predictions of the modulation approach in the strong localization hopping transport regime. The hypothesis of a second order phase transition or percolation transition at a density below that of the MIT is thus reinforced.

**Keywords:** 1/f noise, two-dimensional hole systems, GaAs heterojunction, percolation, hopping.
**PACS:** 71.30.+h, 71.27.+a, 72.70.+m, 73.21.Fg

## INTRODUCTION

The physical nature of electronic systems changes drastically when the density of carriers is reduced. At large densities, the delocalized independent quasi-particle description (Fermi liquid, FL) prevails. At very low density and weak disorder, the ground state of the system is believed to be the Wigner crystal [1,2]. If the disorder is large and the interactions remain weak, decreasing the density leads to the strong localization (SL) regime for independent particles (Anderson insulator) [3]. The challenge for experimentalists is to investigate the strongly correlated systems which should appear as the density is decreased when the disorder is so low that the effect of the interactions should play a major role. Two-dimensional electron or hole systems (2DES or 2DHS) at low temperature offer the possibility of studying such systems as the density is simply tuned by applying a voltage to a metallic gate. The very high mobilities reached in presently available "clean" samples guarantees levels of disorder low enough for such new physics to appear instead of the classical crossover from FL to SL. The ratio which "measures" the magnitude of the interactions between carriers is $r_s = E_{ee}/E_F \propto m^*/p_s^{1/2}$, where $E_{ee}$ and $E_F$ are the interaction and Fermi energies, $m^*$ the effective mass of the carriers, and $p_s$ their areal density. The observation of a metallic behavior for $4 < r_s < 36$, in 2DES or 2DHS in high mobility silicon metal-oxide-semiconductor field effect transistors (Si-MOSFETs) and in GaAs heterostructures has raised the possibility of a new metallic phase due to the interactions [4-9], in contradiction with the scaling theory of localization for independent particles [3]. The metallic behavior is defined by a decrease of the resistivity $\rho$ for decreasing temperature $T$, for $p_s > p_c$ where $p_c$ is a critical density. When $p_s < p_c$, an insulating behavior occurs (d$\rho$/d$T$ < 0). Explanations of the metallic behavior by corrections to the standard independent particle picture have been put forward [4-9], but the question of the physical nature of the system at large $r_s$ remains open. Several scenarios of the physics at large $r_s$ in clean samples have been proposed. The system may freeze into a glass instead of crystallizing, as found in Si-MOSFETs [10-11]. In GaAs 2DHS, local electrostatic studies [12-13] and transport measurements in a parallel magnetic field [14] suggest the coexistence of two phases. Several calculations predict a spatial separation of a low and a high density phase [15-17], and the transport properties could be due to the percolation of the conducting phase through the insulating one. More generally, the percolation scenario has been put forward by several authors [17-22]. Experimentally, scaling laws observed on the resistance [18,19-22] and its fluctuations [21,22] of 2DHS and 2DES in GaAs favor the percolation scenario. While the "metallic" phase has been widely studied, only few experiments have been performed in the low density insulating phase. In the present study, we use $1/f$ noise as a tool to investigate the low-density regime in $p$-GaAs.

## EXPERIMENTAL METHOD

Our 2DHS are created in Si modulation doped (311)A high mobility GaAs quantum wells. The metallic gate used to change the density is evaporated onto a 1 μm thick insulating polymide film [21,22]. The experiments were carried out on Hall bars 50 μm wide and 300 μm long. The mobility at a density $p_s = 6\times10^{10}$ cm$^{-2}$ and a temperature $T = 100$ mK is $5.5\times10^5$ cm$^2$/V.s, a large value which guarantees the very low disorder. The "clean" nature of the 2DHS is confirmed by the temperature dependence of the resistivity ("metallic" behavior: resistivity increases by a factor of almost 2 as $T$ increases [21,22]), and the well defined plateaus and oscillations of the Hall and Shubnikov-de-Haas curves at low density [22]. We performed transport and resistance noise measurements at densities ranging from about $1.0\times10^{10}$ cm$^{-2}$ to $1.6\times10^{10}$ cm$^{-2}$ and temperatures from 50 to 800 mK. Special attention has been paid to minimize parasitic contributions to the noise [22]. Two measurement methods were used: the voltage and its time fluctuations were measured for a fixed injected current $I$, or the current and its time fluctuations were measured, for a fixed applied voltage $V$. Both DC and AC techniques were used. The frequency interval in which the spectra were recorded is typically 0.01 Hz < $f$ < 3 Hz. The final result is the normalized resistance noise power $S_R/R^2$, $R$ being the resistance of the 2DHS in the ohmic region. In the fixed $I$ method, the voltage noise power spectrum $S_V$ was obtained by using the cross-correlation technique [21,22] for the two voltage noise signals measured on opposite sides of a Hall bar to suppress the contribution of the noise due to the contacts, leads and preamplifiers. $[S_V(I) - S_V(0)]/I^2$ did not depend on $I$, and $S_R = [S_V(I) - S_V(0)]/I^2$. In the fixed $V$ method, the current noise power spectrum $S_I$ was obtained by using the cross-correlation technique for the two voltage signals measured on two resistors in series with the sample. We verified that possible spurious noise sources (e.g. fluctuations of *I*, *T*, gate voltage $V_G$, etc.) did not contribute to $S_R$ [21,22]. Fig. 1 shows that the fixed I or V methods give the same result.

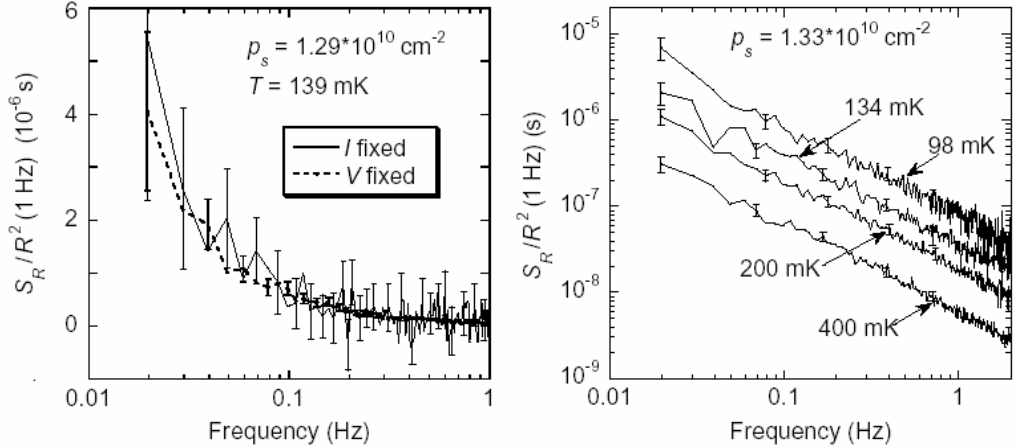

**FIGURE 1.** Left: an example of comparison between the two spectra obtained by using the fixed *I* and fixed *V* method. Right: Typical spectra obtained at four temperatures for $p_s=1.33\times10^{10}$ cm$^{-2}$.

## EXPERIMENTAL RESULTS

The transport measurements give an insulating $\rho(T)$ dependence ($d\rho/dT < 0$) for densities below $p_c \approx 1.46\times10^{10}$ cm$^{-2}$. For low densities and temperatures, the divergence of the resistivity when *T* goes to zero is described by an activated law [21,22]. Fig. 2 gives the noise power for $0.81\times p_c < p_s < 0.96\times p_c$. $S_R/R^2$ increases strongly when $p_s$ or *T* decrease. The slope of the *T* dependence increases when $p_s$ decreases, thus the trend already observed in Ref. [21] for $0.95\times p_c < p_s < p_c$ continues at lower densities. In Ref. [21], we found that the noise scales with the resistance, which suggests a second order phase transition such as a percolation transition at a critical density $p^* < p_c$. The alternative "standard" description consists in a FL-SL crossover, the transport laws in the SL regime being hopping of carriers between localized sites. Let us examine whether our noise results at low density agree or not with the hopping picture. Experimentally, noise data in the 2D SL regime have been successfully analyzed by Pokrovskii et al. [23], using the modulation approach of Kozub [24]. In this model, $S_R/R^2 \sim 1/N\,(\delta\rho_i/\rho_i)^2$ where *N* is the number of bonds in the Miller-Abrahams cluster, $\rho_i$ the resistance of the critical hop, and $\delta\rho_i$ its variation due to the electrostatic influence of fluctuating charges outside the cluster. As our transport law indicates simple activation, we would expect nearest neighbor hopping (NNH) in the SL regime. In NNH, *N* does not depend on the temperature [23,24]. We are thus led to attribute the *T* dependence of $S_R/R^2$ to the $\delta\rho_i/\rho_i$ term. Its temperature dependence has been calculated in Ref. [23]: for a size of the dipole modulator larger (resp. smaller) than the distance between the modulator and the critical hop, it is proportional to $T^{-1/3}$ (resp. $T^{1/9}$). This is far from what is shown in Fig. 2 where the power-like dependences indicate *T* exponents less than -1. We conclude that the *T* dependence of the noise contradicts the hopping between localized sites at least in the modulation approach.

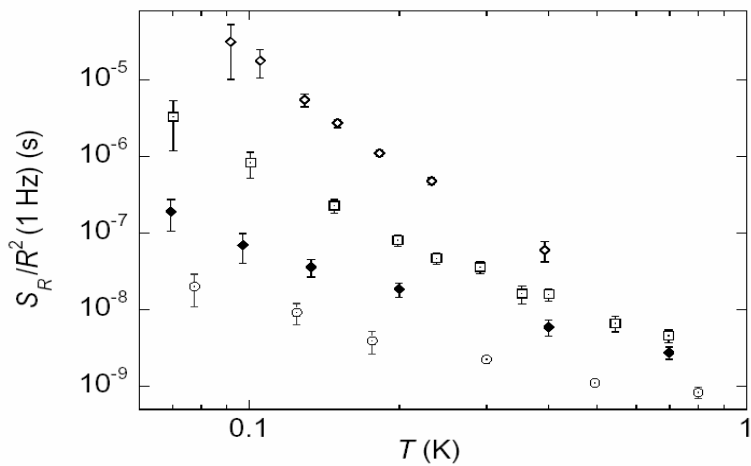

**FIGURE 2.** Relative resistance noise power at 1 Hz as a function of the temperature for four densities: $p_s=1.40\times10^{10}$ cm$^{-2}$ (open circles), $1.33\times10^{10}$ cm$^{-2}$ (closed diamonds), $1.28\times10^{10}$ cm$^{-2}$ (open squares), $1.18\times10^{10}$ cm$^{-2}$ (open diamonds).